\documentstyle[aps,pra,twocolumn]{revtex}
\begin{document}
%\draft

\newcommand{\pp}[1]{\phantom{#1}}
\newcommand{\be}{\begin{eqnarray}}
\newcommand{\ee}{\end{eqnarray}}
\newcommand{\ve}{\varepsilon}
\newcommand{\Tr}{{\rm Tr\,}}
\newtheorem{th}{Theorem}
\newtheorem{lem}[th]{Lemma}

%@twocolumnfalse\endcsname

\title{
Lie-Nambu and beyond
}
\author{Marek Czachor}
\address{
Katedra Fizyki Teoretycznej i Metod Matematycznych\\
Politechnika Gda\'{n}ska,
ul. Narutowicza 11/12, 80-952 Gda\'{n}sk, Poland\\
E-mail: mczachor@sunrise.pg.gda.pl\\
{\rm ***\\
 An extended version of the talk given in July 1997 in
Peyresq, France}
}
\maketitle
\begin{abstract}
Linear quantum mechanics can be regarded as a particular example
of a nonlinear Nambu-type theory. Some elements of
this approach are presented. 
\end{abstract}

\narrowtext
\section{Introduction}

At the moment there is no single experimental data suggesting that
states of a quantum system can evolve in a fundamentally
nonlinear way. 
On the other hand all impossibility theorems stating
that such a nonlinearity is in principle impossible 
have not survived a detailed analysis.
It is therefore possible that the status of quantum linearity is similar
to this of geometrical linearity from before the invention of general 
relativity. 

The multiple-bracket
dynamics described in this paper arose from a search for a
consistent embedding of linear quantum mechanics into a more
general theory where the assumption of linearity could be
dropped. The formalism is essentially
based on density matrices and not on wave functions. 
A density matrix plays here a role of a fundamental field and
should not be regarded as a mixture of classical and quantum
probabilities. A departure point for the discussed
generalization is the observation that density matrices of
ordinary quantum mechanics satisfy an equation of a Lie-Nambu
type. 

The layout of the paper is as follows. Sec. II relates
the work to the earlier ones by Nambu~(1973), and Bia{\l}ynicki-Birula
and Morrison~(1991). Some formal tools are introduced in
Sec. III, IV and V. In Sec. VI a $(2n+1)$-bracket is introduced and
some of its general properties are proved. The bracket differs
from the so-called generalized Nambu, generalized Poisson, or
generalized Nambu-Poisson 
brackets discussed in the literature (Bayen and Flato, 1975; Cohen, 1975;
Takhtajan, 1994; Gautheron, 1996; Chatterjee and Takhtajan,
1996;  de Azc\'arraga et al., 1996; Ditto and Flato, 1997;
Ib\'anez et al., 1997; Ditto et al., 1997; Kanatchikov, 1997). 
The particular case of a 3-bracket dynamics is
discussed in Sec. VII where the notion of a Lie-Nambu duality is
introduced and properties of solutions of the 3-bracket equations
are discussed. The 5-bracket dynamics is briefly discussed in
Sec. VIII. Sec. IX is devoted to the question of $N$-particle
extensions of a nonlinear Lie-Poisson dynamics. The notion of
complete separability is discussed and examples of completely
separable equations are given. The results of this section
contradict the popular belief that all nonlinear extensions of
quantum mechanics lead to faster-than-light phenomena. Also the
question of complete positivity of solutions is shortly
discussed in this Section. Sec. X is devoted to the problem of
separability of the dual Poisson dynamics and it is shown that a
surprising nonlocal phenomenon occurs. In Sec. XI a
possible link between our formalism and the
problem of quantization of classical Nambu dynamics is discussed.

\section{Lie-Poisson as a Lie-Nambu}

The origin of this work goes back to two papers where, in
completely different contexts, a notion of a triple bracket was
introduced. 

\subsection{Nambu 1973: Euler equations}

The Euler equations for a rotating rigid body are 
\be
\dot J_k 
&=& \epsilon_{kbc}(J_b/I_b)J_c\\
&=& J_a\epsilon{^a}_{bc}
\frac{\partial J_k}{\partial J_a}
\frac{\partial H}{\partial J_b}=\{J_k,H\}\label{LP}\\
&=& \epsilon_{abc}
\frac{\partial J_k}{\partial J_a}
\frac{\partial H}{\partial J_b}
\frac{\partial S}{\partial J_c}
=\{J_k,H,S\}.\label{LN}
\ee
Here $\bbox J$ is an angular momentum, $I_k$ a component of a
moment of inertia, $H=\frac{J_1^2}{2I_1}+\frac{J_2^2}{2I_2}
+\frac{J_3^2}{2I_3}$ rotational energy, and $S=\frac{1}{2}\bbox
J^2$. The totally antisymmetric tensor $\epsilon_{abc}$ can be
regarded either as a 3-dimensional volume form, or as structure 
constants of $so(3)$. The Lie algebra $so(3)$ enters the
equations also via $S$ since $\bbox J^2$ is a second order Casimir
invariant of this algebra. The form (\ref{LP}) defines a Poisson
bracket. The triple bracket defined by (\ref{LN}) is
nowadays called the Nambu bracket and was introduced in Nambu~(1973).

The Poisson bracket (\ref{LP}) is a particular case of the
so-called Lie-Poisson bracket which differs from (\ref{LP}) by the presence
of structure constants $c{^a}_{bc}$ of some Lie algebra instead of
$\epsilon{^a}_{bc}$ characteristic of $so(3)$. 
It is natural to think of the Nambu bracket as a particular
case of 
\be
\{A,B,S\}=c_{abc}
\frac{\partial A}{\partial x_a}
\frac{\partial B}{\partial x_b}
\frac{\partial S}{\partial x_c}\label{cLN}
\ee
where $S=\frac{1}{2}g^{ab}x_ax_b$ is a second order Casimir
invariant of an appropriate Lie algebra. Such brackets could be
called the Lie-Nambu ones and are quite natural in the context
of generalizations of a Lie-Poisson dynamics. It is surprising
that this kind of generalization of 
Hamiltonian dynamics has not been
considered so far in the theory of classical dynamical systems (Ratiu, n.d.). 
One of the reasons seems to be the
fact that for general Lie algebras the bracket does not satisfy
the so called fundamental identity (Chatterjee and Takhtajan,
1996). We shall return to this question in Sec. VII. 

The generalization of (\ref{LN}) which was extensively
investigated in the literature under the name of a ``generalized Nambu
dynamics" goes in another direction
(Bayen and Flato, 1975;
Takhtajan, 1994; Gautheron, 1996; 
Chatterjee and Takhtajan, 1996).
One treats the $\epsilon$ not as structure constants
but as a volume form. From this perspective it is natural to
consider 
\be
\{A_1,\dots ,A_n\}=\epsilon_{a_1\dots a_n}
\frac{\partial A_1}{\partial x_{a_1}}
\dots
\frac{\partial A_n}{\partial x_{a_n}}.
\ee
The parameter $n$ is a dimension of the state space. 
It is not clear how to extend this type of description to
infinite-dimensional spaces. 

\subsection{Bia{\l}ynicki-Birula--Morrison 1991: Liouville--von
Neumann equation}

The observation that the Liouville--von Neumann
equation for a Wigner function can be written as a Lie-Nambu
equation with nontrivial structure constants 
is due to Bia{\l}ynicki-Birula and Morrison (1991). 
Below,
instead of the Wigner function which is defined in
terms of position-momentum coordinates, we shall 
stick to the more symmetric position-position representation.
This will lead to a specific form of structure constants
whose symmetry properties will be essential for further
generalizations (Czachor, 1997a; Czachor and Kuna, 1997a). 

The density matrix in position representation is denoted by
$\rho(a,a')=:\rho_a$, where we use $a$ and $a'$ instead of  more
typical $x$ and $x'$, and the lower composite index is
introduced for brievity. The kinetic energy is represented by
the kernel 
\be
\int dy\,K(a,y)\rho(y,a')=\frac{-\Delta_{a}}{2m}\rho(a,a'),
\ee
and the Hamiltonian operator by 
\be
H(a',a)&=&
K(a',a) + V(a)\delta(a-a').
\ee
It is easy to check that the following equation 
\be
{}&{}&i\partial_t\rho_a =
\int db  db' dc  dc'\nonumber\\
&{}&\times\big(
\underbrace{
\delta(a-b')\delta(b-c')\delta(c-a')
-
\delta(a-c')\delta(b-a')\delta(c-b')
}_{\Omega_{abc}}\big)\nonumber\\
&{}&\times\underbrace{H(b',b)}_{H^b} \underbrace{\rho(c',c)}_{\rho^c}
=
\Omega_{abc}H^b\rho^c\label{LvN}
\ee
is equivalent to the Liouville--von Neumann one. The form
(\ref{LvN}) simultaneously illustrates the use of composite
indices and the summation convention. Notice that  the
composite indices are in their lower or upper position, and the
transition between the two is given by the metric tensor 
$g^{ab}$ working as follows
\be
g^{ab}\rho_b
&=&
\int db db'\delta(a-b')\delta(b-a')\rho(b,b')=\rho(a',a)=\rho^a.\nonumber
\ee
So if $\rho_a=\rho(a,a')$ then $\rho^a=\rho(a',a)$. 
Although the latter formula may seem somewhat artificial and was
not used by Bia{\l}ynicki-Birula and Morrison, it will
prove extremely useful when we arrive at various generalizations. 
The distributions $\Omega_{abc}$ are structure constants of an
infinte-dimensional Lie algebra, which can be checked
by raising $a$ with the help of $g^{ab}$ and verifying the standard
properties. One should be aware of the fact that $g^{ab}$ is
{\it not\/} the Cartan-Killing metric (which does not exist
in this case). Writing $g^{ab}\rho_a\rho_b=\Tr(\hat \rho^2)=:C_2=:2S$ one
recognizes that $g^{ab}$ is a kernel form of the Hilbert-Schmidt   
metric. Let now $H(\hat \rho)=\Tr \hat H\hat \rho$. Taking into
account that $H^a=\delta H/\delta \rho_a$ and $\rho^a=\delta
S/\delta \rho_a$ ($\delta /\delta \rho_a$ is a functional
derivative) we can write the Liouville--von Neumann equation in
the Bia{\l}ynicki-Birula--Morrison form as
\be
i \dot \rho_k
&=& 
\rho_a\Omega{^a}_{bc}\frac{\delta \rho_k}{\delta \rho_b}
\frac{\delta H}{\delta \rho_c}
=\{\rho_k,H\}\label{qLP}\\ 
&=& 
\Omega_{abc}\frac{\delta \rho_k}{\delta \rho_a}
\frac{\delta H}{\delta \rho_b}
\frac{\delta S}{\delta \rho_c}
=\{\rho_k,H,S\}.\label{qLN} 
\ee
Although the relationship of (\ref{qLP}) and (\ref{qLN}) to 
(\ref{LP}) and (\ref{LN}) is obvious, it requires a few comments.
First of all, the equations are Lie-Poisson and {\it
Lie\/}-Nambu and not the ``generalized Nambu" in the sense of the
previous subsection.  The Lie algebra and the space of states 
are both infinite-dimensional. 
$H$ is an average energy and $C_2=2S$ is a Casimir invariant
called an ``entropy". (Actually, it {\it is\/}, up to constants,
a 2-entropy of 
Dar\'oczy (1970) or Tsallis (1988), and is closely
related to R\'enyi's $\alpha$-entropies (R\'enyi, 1960, 1961).)  

\section{Digression on pure states}

There is one more formal prerequisite we need before we get
further. The Liouville--von Neumann equation for mixed states has its
roots in the pure-state Schr\"odinger equation. It turns out
that the same is true of the structure constants $\Omega_{abc}$.
Let us switch now one level higher and instead of speaking about a
nonrelativistic, spin-0 Schr\"odinger equation consider 
general Hamilton equations on a separable Hilbert space:
\be
i \omega^{\alpha\alpha'}\dot \psi_\alpha
=
\frac{\delta H}{\delta\bar \psi_{\alpha'}},
\quad
-i \omega^{\alpha\alpha'}\dot {\bar \psi}_{\alpha'}
=
\frac{\delta H}{\delta\psi_{\alpha}}.
\label{Ham}
\ee
If $\psi_\alpha=\psi(a)$,
$\omega^{\alpha\alpha'}=\delta(a-a')$, $H=\langle\psi|\hat H
|\psi\rangle$, and an obvious summation/integration  convention is applied,
(\ref{Ham}) is equivalent to the Schr\"odinger equation
(Chernoff and Marsden, 1974). $\omega^{\alpha\alpha'}$ is a symplectic form
in the complex coordinates ``$\psi=q+ip$". An explicit form of
$\omega^{\alpha\alpha'}$ varies from representation to
representation and is different for, say, the Dirac equation, or a
nonrelativistic particle with spin. It is important however that the
form of the Hamilton equations (\ref{Ham}) is always the same
(although the dot at its LHS may have different meanings as well, cf.
Czachor (1997a), Czachor and Kuna (1997a)) and that
$\omega^{\alpha\alpha'}\psi_\alpha
\bar\phi_{\alpha'}=\langle\phi| \psi\rangle$.  (\ref{Ham}) can
be written in a form involving a Poisson tensor
\be
i \dot \psi_\alpha
=
I_{\alpha\alpha'}
\frac{\delta H}{\delta\bar \psi_{\alpha'}},
\quad
-i \dot {\bar \psi}_{\alpha'}
=
I_{\alpha\alpha'}
\frac{\delta H}{\delta\psi_{\alpha}}.\label{Ham'}
\ee
A pure-state density matrix is $\rho_a=\rho_{\alpha\alpha'}=
\psi_\alpha\bar \psi_{\alpha'}$ and $\omega^{\alpha\alpha'}
\rho_{\alpha\alpha'}=\omega^a\rho_a=\Tr \hat \rho$.
A pure-state Poisson bracket corresponding to (\ref{Ham'}) and
its complex conjugated equation is 
\be
\{A,B\} &=& I_a\frac{\delta A}{\delta\psi_\alpha}
\frac{\delta B}{\delta\bar \psi_{\alpha'}}
-
(A\leftrightarrow B)
\\
&=&
\rho_a\Omega{^a}_{bc}\frac{\delta A}{\delta \rho_b}
\frac{\delta B}{\delta \rho_c},
\ee
which holds for all functions $A(\rho_a)=A(\psi_\alpha\bar \psi_{\alpha'})$
and $B(\rho_a)=B(\psi_\alpha\bar \psi_{\alpha'})$. 
The structure constants are 
\begin{eqnarray}
\Omega{^a}_{bc}&=&\delta_{\beta'}{^{\alpha'}}\delta_{\gamma}{^{\alpha}}
I_{\beta\gamma'}-
\delta_{\gamma'}{^{\alpha'}}\delta_{\beta}{^{\alpha}}
I_{\gamma\beta'}\\
\Omega_{abc}&=&I_{\alpha\beta'}I_{\beta\gamma'}
I_{\gamma\alpha'}-
I_{\alpha\gamma'}I_{\beta\alpha'}
I_{\gamma\beta'}\\
\Omega^{abc}&=&-\omega^{\alpha\beta'}\omega^{\beta\gamma'}
\omega^{\gamma\alpha'}+
\omega^{\alpha\gamma'}\omega^{\beta\alpha'}
\omega^{\gamma\beta'},
\end{eqnarray}
where the deltas are defined by
\begin{eqnarray}
I_{\alpha\beta'}\omega^{\alpha\alpha'}&=&\delta_{\beta'}{^{\alpha'}}\\
I_{\beta\alpha'}\omega^{\alpha\alpha'}&=&\delta_{\beta}{^{\alpha}}.
\end{eqnarray}
The metric tensor that raises and lowers the composite indices
is $g^{ab}=\omega^{\alpha\beta'}\omega^{\beta\alpha'}$ and 
$g_{ab}=I_{\alpha\beta'}I_{\beta\alpha'}$.
The reader may check that we obtain the
Bia{\l}ynicki-Birula--Morrison formulas if we replace $\omega$'s
and $I$'s by the Dirac deltas. 

\section{Higher order ``metric" tensors}

In this Section we introduce several technical 
results which will turn our abstract composite index language
into a practical tool.

We have seen that $\Tr(\hat \rho^2)=g^{ab}\rho_a\rho_b$. It is
useful to introduce higher order tensors 
satisfying $\Tr(\hat \rho^n)=g^{a_1\dots a_n}\rho_{a_1}\dots
\rho_{a_n}$. Define 
\be
g^{a_1\dots a_n}
&=&
\omega^{\alpha_1\alpha'_n}
\omega^{\alpha_2\alpha'_1}\omega^{\alpha_3\alpha'_2}\dots
\omega^{\alpha_{n-1}\alpha'_{n-2}}
\omega^{\alpha_n\alpha'_{n-1}},\label{m1'}\\
G_{a_1\dots a_n}
&=&
I_{\alpha_1\alpha'_n}
I_{\alpha_2\alpha'_1}I_{\alpha_3\alpha'_2}\dots
I_{\alpha_{n-1}\alpha'_{n-2}}
I_{\alpha_n\alpha'_{n-1}}.\label{m2'}
\ee
If we lower the indices in (\ref{m1'}) we see that, somewhat
counter-intuitively, $g$ does not go directly into $G$ (although
$g_{ab}\omega^b=I_a$!) but 
\be
g_{a_1b_1}\dots g_{a_nb_n} g^{b_1\dots b_n}
=g_{a_1\dots a_n}=
G_{a_na_{n-1}\dots a_1}.
\ee
So define a $*$-operation which reverses the order of indices:
$^{**}={\rm id}$, ${g^*}_a=g_a$, 
${g^*}_{ab}=g_{ba}=g_{ab}$, 
and 
\be
{g^*}_{a_1\dots a_n}=G_{a_1\dots a_n}=g_{a_na_{n-1}\dots a_1}
\ee
and similarly with the upper indices. 
Essential for further calculations are the following properties.

\medskip
\noindent
a) Cyclicity
\be
g_{a_1\dots a_nb_1\dots b_m}&=&g_{b_1\dots b_ma_1\dots a_n},\\
{g^*}_{a_1\dots a_n b_1\dots b_m} &=& {g^*}_{b_1\dots b_ma_1\dots a_n}.
\ee
b) ``Cut-and-glue" 
\be
{g^*}_{a_1\dots a_nx}{g^*}{^x}_{a_{n+1}\dots a_{n+m}}
&=&
{^*g}_{a_1\dots a_{n+m}}\\
g^{a_1\dots a_nx}g{_x}^{a_{n+1}\dots a_{n+m}}
&=&
g^{a_1\dots a_{n+m}}.
\ee
c) ``Annihilation" 
\be
I_{x}g^{a_1\dots a_{k}xa_{k+1}\dots a_{n+m}}
&=&
g^{a_1\dots a_{k}a_{k+1}\dots a_{n+m}}\\
\omega^{x}g_{a_1\dots a_{k}xa_{k+1}\dots a_{n+m}}
&=&
g_{a_1\dots a_{k}a_{k+1}\dots a_{n+m}}
\ee
d) ``Drag-and-drop"
\be
g^{b_1\dots b_l}{_{x}}g^{a_1\dots a_{k}xa_{k+1}\dots a_{n+m}}
&=&
g^{a_1\dots a_{k}b_1\dots b_l a_{k+1}\dots a_{n+m}}\\
g_{b_1\dots b_l}{^{x}}g_{a_1\dots a_{k}xa_{k+1}\dots a_{n+m}}
&=&
g_{a_1\dots a_{k}b_1\dots b_l a_{k+1}\dots a_{n+m}}.
\ee
It is practical to accept the rule stating that complex
conjugation interchanges primed and unprimed indices. Assuming this
we can define symmetric operators $\hat A$ as those whose
kernels satisfy
$\overline{A_{\alpha\beta'}}=A_{\beta\alpha'}$.
We find also  that 
\be
\overline{g_{a_1\dots a_{n}}}={g^*}_{a_1\dots a_{n}}.
\ee
As a consequence 
\be
\overline{g_{a_1\dots a_{n}}A_1^{a_1}\dots
A_n^{a_{n}} }=
g_{a_1\dots a_{n}}A_{n}^{a_1}\dots
A_1^{a_{n}}
\ee
which is an abstract-index version of the well known rule 
\be
\overline{\Tr(\hat A_1\dots
\hat A_n)}=
\Tr(\hat A_n\dots
\hat A_1)
\ee
valid for symmetric operators. In order to translate the
abstract-index formulas into more standard operator ones one
uses the following correspondence:
\be
(\hat A_1\dots
\hat A_n)_a=
g_{aa_1\dots a_n}A_1^{a_1}\dots A_n^{a_n}.
\ee

\section{Structure constants revisited}

A Lie-Nambu 3-bracket written in the form (\ref{cLN}) is based
on a totally antisymmetric 3-index tensor. Obviously, the tensor has
3-indices for all Lie algebras and for this reason it is not
immediately clear whether a generalization of (\ref{cLN}) to a
``generalized Nambu" $n$-bracket is possible. 
On the other hand, the structure constants occuring in (\ref{qLN})
have a rich structure and it turns out there exists a natural
generalization of (\ref{qLN}). 

To begin with let us note that 
\be
\Omega_{abc}&=&g_{abc}-g_{acb}=2!g_{a[bc]}=2!g_{[abc]},\\
\Omega^{abc}&=&g^{abc}-g^{acb}=2!g^{a[bc]}=2!g^{[abc]},
\ee
where $[\dots ]$ denotes an antisymmetrization.

Consider 
\be
\Omega_{a_1\dots a_n}=(n-1)!g_{[a_1\dots a_n]}.
\ee
{\bf Lemma~1.} 
\be
g_{[a_1\dots a_{2m}]}&=&0\label{L1.1}\\
g_{[xa_1\dots a_{2m}]}&=&g_{x[a_1\dots a_{2m}]},\label{L1.2}\\
\omega^x\Omega_{a_1\dots x\dots a_n}&=&0.\label{L1.3}
\ee
{\it Proof\/}: (a) Eq.~(\ref{L1.1}).
\be
g_{[a_1\dots a_{n}]}=g_{[a_2\dots a_{n}a_1]}=
(-1)^{n-1}g_{[a_1\dots a_{n}]},\nonumber
\ee
where the cyclicity and total antisymmetry were used. The
expression vanishes for even $n$. (b) Eq.~(\ref{L1.2}). Assume $n=2m$.
\be
{}&{}&g_{[xa_1\dots a_n]}\nonumber\\
&{}&=
\frac{1}{n+1}
\Big(
g_{x[a_1\dots a_n]}+\dots +(-1)^kg_{[a_1\dots a_k|x|a_{k+1}\dots
a_n]}+\dots\nonumber\\
&{}&\phantom{
=
\frac{1}{n+1}
\Big(
g_{x[a_1\dots a_n]}+
}\dots +
(-1)^ng_{[a_1\dots a_{n}]x}\Big)\nonumber\\ 
&{}&=
\frac{1}{n+1}\Big(
g_{x[a_1\dots a_n]}+\dots +(-1)^kg_{x[a_{k+1}\dots
a_na_1\dots a_k]}+\dots\nonumber\\
&{}&\phantom{
=
\frac{1}{n+1}
\Big(
g_{x[a_1\dots a_n]}+
}\dots + 
g_{x[a_1\dots a_{n}]}\Big)\nonumber\\
&{}&=
\frac{1}{n+1}
\Big(
g_{x[a_1\dots a_n]}+\dots +(-1)^{k+(n-k)k}g_{x[a_1\dots a_n]}
+\dots \nonumber\\
&{}&\phantom{
=
\frac{1}{n+1}
\Big(
g_{x[a_1\dots a_n]}+
}\dots+ 
g_{x[a_1\dots a_{n}]}\Big)\nonumber\\
&{}&=
g_{x[a_1\dots a_{n}]}\nonumber
\ee
where we have used the cyclicity and the fact that 
$(n-k+1)k$ is even for any $k$ if $n$ is even. 
(c) Eq.~(\ref{L1.3}). It is
sufficient to note that the annihilation property together with
(\ref{L1.1}) and (\ref{L1.2}) imply 
\be
\omega^xg_{[xa_1\dots a_{2m}]}=\omega^xg_{x[a_1\dots a_{2m}]}=
g_{[a_1\dots a_{2m}]}=0.\nonumber
\ee
$\Box$

\section{Generalized Lie-Nambu brackets}

We define the generalized Lie-Nambu bracket for $n=2m+1$ by
\be
\{A_1,\dots,A_n\}
=
\Omega_{a_1\dots a_n}
\frac{\delta A_1}{\delta \rho_{a_1}}\dots
\frac{\delta A_n}{\delta \rho_{a_n}}.\label{genLN}
\ee
Let $C_k=g^{a_1\dots a_k}\rho_{a_1}\dots\rho_{a_k}$. 

\medskip\noindent
{\bf Theorem 1.}
\be
\{C_{k_1},\dots,C_{k_{(n+1)/2}},\, \cdot\,,\dots,\,\cdot\,\}=0.
\ee
{\it Proof\/}:
Let us begin with the following remark. Assume a tensor 
$F_{\dots abc\dots}$ has the ``drag-and-drop" property
\be
F_{\dots abc\dots}g{^c}{_{c_1c_2}}=F_{\dots abc_1c_2\dots}\nonumber
\ee
and consider
\be
{}&{}&
F_{\dots[abc]\dots}\rho^ag{^c}{_{c_1c_2}}\rho^{c_1}\rho^{c_2}\nonumber\\
&{}&=
\frac{1}{6}
\Big(
F_{\dots abc\dots}
+
F_{\dots bca\dots}
+
F_{\dots cab\dots}\nonumber\\
&{}&\phantom{
=
\frac{1}{6}
\Big(
}
-
F_{\dots acb\dots}
-
F_{\dots cba\dots}
-
F_{\dots bac\dots}
\Big)
\rho^ag{^c}{_{c_1c_2}}\rho^{c_1}\rho^{c_2}\label{6terms}\\
&{}&=
\frac{1}{6}
\Big(
F_{\dots ab{c_1c_2}\dots}
+
F_{\dots b{c_1c_2}a\dots}
+
F_{\dots {c_1c_2}ab\dots}\nonumber\\
&{}&\phantom{
=
\frac{1}{6}
\Big(
}
-
F_{\dots a{c_1c_2}b\dots}
-
F_{\dots {c_1c_2}ba\dots}
-
F_{\dots ba{c_1c_2}\dots}
\Big)
\rho^a\rho^{c_1}\rho^{c_2}\nonumber\\
&{}&=
\frac{1}{6}
\Big(
F_{\dots ab{c_1c_2}\dots}
-
F_{\dots {c_1c_2}ba\dots}
\Big)
\rho^a\rho^{c_1}\rho^{c_2}\nonumber
\ee
which in general does not vanish. We can see therefore that for
expressions such as (\ref{6terms}) to vanish, it is sufficient
to have in each term of the decompositition (\ref{6terms})
at least two transvected indices which are not seperated by a
non-transvected one. 
With this observation in mind consider 
\be
{}&{}&g_{x[a_1\dots a_Nc_1\dots c_M]}
{g^{a_1}}_{a_1^1\dots a_1^{K_1}}\dots
{g^{a_N}}_{a_N^1\dots a_N^{K_N}}\nonumber\\
&{}&\phantom{xxxxxxxxxxxxxxx}\rho^{a_1^1}\dots \rho^{a_1^{K_1}}\dots
\rho^{a_N^1}\dots \rho^{a_N^{K_N}}.\label{b[...]}
\ee
Expanding (\ref{b[...]}) we will obtain a sum involving expressions
$g_{x\dots ij\dots}$ and $g_{x\dots ji\dots}$ entering with
opposite signs. If $N$ and $M$ will be chosen in a way
guaranteeing that for any such term there exists at least a pair
$(a_k,a_l)$ of indices that are not separated by some $c_r$ then
(\ref{b[...]}) vanishes on the basis of the preceding argument.
We know that $N+M$ is an even number. Therefore the greatest $N$ that allows
for a separation of any two $a_k$'s is $N=(N+M)/2=M=(n-1)/2$.
To complete the proof it is sufficient to note that (\ref{b[...]}) equals
\be
\frac{1}{k_1}\dots \frac{1}{k_N}
\{C_{k_1},\dots ,
C_{k_N},\,\cdot\,,\dots,\,\cdot\,\},
\ee
where $k_j=K_j+1$.$\Box$

\medskip\noindent
{\it Remarks\/}: 
(1) Total antisymmetry of (\ref{genLN}) guarantees that
$\{C_k,C_k,\dots\}=0$. Moreover (\ref{L1.3}) leads to $\{C_1,\dots\}=0$. 
Since $C_1=\Tr\hat \rho$ the dynamics generated by such brackets
is trace preserving. 
(2) For any $k$, $l$ $\{C_k,C_l,\dots\}\big|_{\rm
pure\,states}=0$. This follows from 
\be
g^{ab_1\dots b_n}\rho_{b_1}\dots \rho_{b_n}|_{\rm
pure\,states}=\rho^aC_1^{n-1}|_{\rm
pure\,states}.
\ee
(3) Theorem~1 was proved for $n=3$ in Czachor~(1997a). 

\medskip\noindent
{\bf Theorem~2.} Let $S_j=S_j(C_1,C_2,\dots)$ be a
differentiable function of $C_k$, $k=1,2,3,\dots$ and $z_n$ 
a complex number.
The dynamics given by 
\be
\dot\rho_a = z_n\{\rho_a,
H_1,\dots,H_{(n-1)/2},S_1,\dots,S_{(n-1)/2}\}\label{eq} 
\ee
conserves $C_k$. $C_k$ are Casimir invariants i.e.
\be
\{C_k, A_1,\dots,A_{(n-1)/2},S_1,\dots,S_{(n-1)/2}\}=0
\ee
for any functions $A_k$. 

\medskip\noindent
{\it Remarks\/}: (1) Theorem~2 is a straightforward consequence of
Theorem~1. (2) The number $z_n$ will be assumed to satisfy $\bar
z_n=-z_n$ (for $n=4m+3$) or 
$\bar z_n=z_n$ (for $n=4m+1$), $m=0,1,2,\dots$. The simplest
choice is therefore either $z_n=-i$, for $n=4m+3$, or $z_n=1$,
for $n=4m+1$ (see the discussion below).  

\section{3-bracket}

The simplest $n=4m+3$ case is $n=3$. The discussion given by
Bia{\l}ynicki-Birula and Morrison 
dealt with {\it linear\/} quantum mechanics. A possibility of
using the 3-bracket dynamics as a departure point for
{\it nonlinear\/} generalizations of quantum mechanics was
described in some detail in Czachor~(1997a). One of the main
motivations for studying the 3-bracket dynamics was a
possibility of introducing nonlinearities only by
generalizations of $S$ and without modifications of $H$. Generalizations via
nonlinear $H$ are interesting and will be discussed in the next sections. 
An important drawback of such Hamiltonian generalizations is
that we have to represent observables by nonlinear operators
which leads to interpretational difficulties. To give an example,
it is not clear which definition of a nonlinear eigenvalue is
physically meaningful, or how to represent higher moments of
experimentally measured random variables if nonlinear operators
are involved (Czachor, 1996a).  
Let us therefore first consider what happens if $H=H^a\rho_a=\Tr
\hat H\hat \rho$ and $S$ is an arbitrary (differentiable) function of the
Casimirs $C_k$. It is easy to see that the dynamics given by a
3-bracket is then linear if and only if $S$ is linear in $C_2$
(Czachor, 1997a). 

\subsection{Lie-Nambu duality}

Is the 3-bracket a Poisson bracket?
The answer to this question reveals an interesting duality which
points into two different generalizations of linear quantum
mechanics. 
To understand the problem define 
$\{A,B\}_X:=\{A,B,X\}$ and check whether the Jacobi identity is
satisfied. Consider
\be
{}&{}&\bigl\{\{A,B\}_X,C\bigr\}_X+\bigl\{\{C,A\}_X,B\bigr\}_X
+\bigl\{\{B,C\}_X,A\bigr\}_X
\nonumber\\
&{}&=
\frac{\delta A}{\delta\rho_{d}}
\frac{\delta B}{\delta\rho_{e}}
\frac{\delta^2 X}{\delta\rho_{a}\delta\rho_{f}}
\frac{\delta C}{\delta\rho_{b}}
\frac{\delta X}{\delta\rho_{c}}
\nonumber\\
&{}&\phantom{xxxxxx}\times\bigl(
\Omega_{def}\Omega_{abc}+\Omega_{bdf}\Omega_{aec}
+\Omega_{ebf}\Omega_{adc}
\bigr).\label{J2}
\ee
The terms involving second derivatives of $A$, $B$ and $C$ drop
out just because of the total antisymmetry of structure
constants. The term involving the second drivative of $X$
vanishes in several cases. For $X=S=g^{ab}\rho_a\rho_b/2$ the
second derivative gives $g^{af}$ and (\ref{J2}) vanishes on the
basis of the structure constants version of the Jacobi identity.
With this choice of $X$ the bracket $\{\,\cdot\,,\,\cdot\,\}_S$
is a Lie-Poisson bracket and the dynamics given by 
\be
\dot \rho_a=-i\{\rho_a,H\}_S=-i\{\rho_a,H,S\}\label{_X}
\ee
is an ordinary Lie-Poisson dynamics. If $H$ is nonlinear the
dynamics corresponds to the nonlinear quantum mechanics in the
Bona-Jordan version (Bona, 1991; Jordan, 1993). 
It can be shown that such brackets satisfy the Jacobi identity
for all $S=S(C_2)$ (Czachor, 1996b).
If $S$ is a function of higher
order Casimirs, say, $S=C_3$, the Jacobi identity does not hold. 
However, rewriting (\ref{_X}) as 
\be
\dot \rho_a=-i\{\rho_a,S\}_{-H}=-i\{\rho_a,S,-H\}\label{_H}
\ee
we obtain a Poisson bracket for any $S$ if $X=-H$ is {\it
linear\/}. It follows that the requirement that {\it
observables\/} are linear leads us, via the Lie-Nambu 3-bracket,
to a dual Poisson structure given by (\ref{_H}). 
This {\it
Lie-Nambu duality\/} $(H,S) \leftrightarrow (S,-H)$ is typical
of all Lie-Nambu theories and is analogous to the canonical
$(q,p) \leftrightarrow (p,-q)$ and electromagnetic $(E,B)
\leftrightarrow (B,-E)$ dualities. 
The duality transformation is a particular case of the duality
rotation
\be
\{\,\cdot,\,H,S\}=\{\,\cdot,\,H\cos\alpha-S\sin\alpha,
H\sin\alpha+S\cos\alpha\}.
\ee
It is noteworthy that the since the $X$-bracket does not in
general satisfy the Jacobi identity, the 3-bracket cannot
satisfy the so-called fundamental identity discussed in
Takhtajan (1994). The example of the Lie-Nambu duality, where
the Jacobi identity simultaneously holds and does not hold
(depending on the viewpoint), clearly
shows that status of such identities is more technical than
fundamental. Each kind of dynamics seems to have its own
fundamental criteria of sensibility. In this work we insist on
positivity of density matrices and lack of faster-than-light
effects. 

Interesting in the context of the duality 
is the pure-state case where $C_n=\langle
\psi|\psi\rangle^n$, $S=S(||\psi||)$, and $H=\langle\psi|\hat
H|\psi\rangle$. The Hamilton equations (\ref{Ham'}) can be
written as 
\be
i \dot \psi_\alpha
=
H_{\alpha\alpha'}
\frac{\delta I}{\delta\bar \psi_{\alpha'}},
\quad
-i \dot {\bar \psi}_{\alpha'}
=
H_{\alpha\alpha'}
\frac{\delta I}{\delta\psi_{\alpha}},\label{Ham''}
\ee
where
$I=\omega^{\alpha\alpha'}\psi_\alpha\bar\psi_{\alpha'}=||\psi||^2$.
If $\psi=\psi_{AB}$ is the electromagnetic
spinor its squared norm $||\psi||^2$ is equal to the classical
energy of the field. Therefore here $H$ is {\it not\/} the
energy although formally it is an expression analogous to the
Dirac Hamiltonian function (Bia{\l}ynicki-Birula, 1996). It is known that the
Hamiltonian formalism based on the energy density $|\psi(x)|^2=\bbox
E(x)^2+\bbox B(x)^2$ corresponds to the Poisson tensor which
involves the differential operator $\vec J\cdot i\vec \nabla=
\vec \nabla \times$ (Bia{\l}ynicki-Birula and
Bia{\l}ynicka-Birula, 1976). On the other hand taking 
$H=\langle\psi|\vec J\cdot i\vec \nabla
|\psi\rangle$ as the Hamiltonian function one gets the Poisson
tensor which involves no differentiations. 
The nonlinear electrodynamics of the Born-Infeld type
(Pleba\'nski, 1970) may be
regarded as a nonlinear generalization within the
$\{\,\cdot\,,S\}_{-H}$ scheme. 

Another manifestation of the duality can be seen in a simpler
case of $N$ harmonic oscillators. The classical energy
$E=\sum(p_k^2+q_k^2)= \sum |p_k+iq_k|^2=\sum |\psi_k|^2$ can be
regarded as a {\it norm\/} squared in the Hilbert space $\bbox
C^N$. It is an easy exercise to rewrite the equations of motion as an
$N$-dimensional Schr\"odinger equation with $\hat H$ being a
diagonal matrix whose eigenvalues are the energies of the
oscillators, but then $H=\langle\psi|\hat
H|\psi\rangle$ is not the energy. 

The above facts suggest an alternative interpretation of the
Poisson structures that occur in quantum mechanics: A
modification of $H$ (say, by interactions) 
can be understood as a deformation of the
Poissonian structure $\{\,\cdot\,,\,\cdot\,\}_{-H}$ of the
manifold of states, and not as a modification of a Hamiltonian
function. Keeping $H$ unchanged but modifying $S$ one
changes a flow on the Poisson manifold but the structure of the
manifold itself is unchanged. 

\subsection{Canonical transformations}

Nonlinear quantum mechanics based on $\{\,\cdot\,,\,\cdot\,\}_S$
uses nonlinear $H$ and $S=S(C_2)$. The canonical transformations
must therefore 
be those that do not change $C_2=\Tr(\hat \rho^2)$ (or $\langle
\psi|\psi\rangle$ for pure states). Such transformations can be
nonlinear and were discussed by Weinberg (1989) and Jordan
(1993). The version 
based on $\{\,\cdot\,,\,\cdot\,\}_{-H}$ leads to canonical
transformations that keep $H=\Tr \hat H\hat \rho$ linear (or 
$\langle\psi|\hat H|\psi\rangle$ bilinear for pure states). The
two classes of transformations are not equivalent. 
It is natural to require that only
those observables which commute with $H$
have to be represented by linear functionals. Various Poissonian
structures that appear in this context may be used also for a
combined quantum-classical description as shown by Jones (1993, 1994)
for the Weinberg type theory.

\subsection{Formal solutions}

Consider first the dynamics with $S=C_{k+1}/(k+1)$. The 3-bracket equation
\be
\dot \rho_a=
-i\{\rho_a,H,S\}=
-i\Omega_{abc}H^bg^{cc_1\dots c_k}\rho_{c_1}\dots 
\rho_{c_k}
\ee
has solutions which can be formally written as 
\be
\rho_a(t)=\sum_{n=0}^\infty
\frac{(-it)^n}{n!}\underbrace{\{\dots\{\{}_n\rho_a(0),H,S\},H,S\}\dots,H,S\}.
\label{szereg}
\ee
Let $\rho_a(0)=\overline{\rho_a(0)}$. In the standard notation
we have $\hat \rho(0)=\hat \rho(0)^*$ and 
\be
\dot{\hat \rho}=-i[\hat H,\hat \rho^k].\label{rho^k}
\ee
For $\hat \rho=\hat \rho^2$ (\ref{rho^k}) is the ordinary linear
Liouville-von Neumann equation.
Asuming $\hat H^*=\hat H$ we find $\dot{\hat \rho}(0)^*=\dot{\hat
\rho}(0)$. In the same way we can prove that 
$d^n\hat \rho/dt^n|_{t=0}=(d^n\hat \rho/dt^n|_{t=0})^*$. It
follows that the formal solution satisfies $\hat \rho(t)=
\hat \rho(t)^*$ if $\hat \rho(0)=\hat \rho(0)^*$. 
The same argument applies to more general $S=S(C_1,C_2,\dots)$. 
(\ref{rho^k}) is interesting in itself even in
finite-dimensional cases where the above argument can be made
more rigorous.
To show that spectrum of self-adjoint 
Hilbert-Schmidt solutions of 
the 3-bracket equations of motion is conserved by the 3-bracket
dynamics  one uses the following 

\bigskip\noindent
{\bf Lemma~2.} Consider a sequence of probabilities
$\{p_k\}_{k=0}^\infty$ and an arbitrary real sequence
$\{a_k\}_{k=0}^\infty$ satisfying for any natural $n$
\be
\sum_{k=0}^\infty p_k^n=\sum_{k=0}^\infty a_k^n.\label{sum}
\ee
Then the two sequences are identical up to permutation.

\bigskip\noindent
{\it Remarks\/}: (1) Let the two sequences represent spectra of
a Hermitian Hilbert-Schmidt solution of the
$n$-bracket equation at $t=0$ and $t\neq 0$ respectively. 
Since $t\mapsto \hat \rho(t)$ is continuous, the spectrum of
$\hat \rho(t)$ is also continuous and, hence, conserved.
The condition 
(\ref{sum}) is implied by conservation of $C_n$. (2) Lemma~2 was
proved in Czachor and Marciniak (1997). (3) 
Such solutions can be interpreted as nonlinearly evolving
density matrices.  The question of their {\it complete\/}
positivity will be discussed below.

\section{5-bracket}

$n=5$ is the simplest (nontrivial) $4m+1$ case. The equation of
motion is ($z_5=1$) 
\be
\dot \rho_a
&=&
\{\rho_a,H_1,H_2,S_1,S_2\}.\label{5}
\ee
Assume that $H_k$ are linear in $\rho$. The simplest 
choice of the other two generators is $S_1=C_2/2$, $S_2=C_3/3$.
Nontrivial 5-bracket equations of motion (\ref{5}) are always nonlinear,
as opposed to the 3-bracket ones which can be linear, and always
vanish on pure states. The RHS of
(\ref{5}) when written in the standard notation involves
an antisymmetrized product of $\hat H_1$, $\hat H_2$, $\hat
\rho$ and $\hat \rho^2$. After a simplification one finds
\be
\dot {\hat \rho}
&=&
\big(
[\hat \rho,\hat H_1]\hat H_2
-
[\hat \rho,\hat H_2]\hat H_1
\big)
\hat \rho^2
\nonumber\\
&\pp =&
+
\hat \rho^2
\big(
\hat H_2[\hat H_1,\hat \rho]
-
\hat H_1[\hat H_2,\hat \rho]
\big)
\nonumber\\
&\pp =&
+
\hat \rho\big(\hat H_2\hat \rho^2\hat H_1
-
\hat H_1\hat \rho^2\hat H_2\big)
\nonumber\\
&\pp =&
+
\big(
\hat H_1\hat \rho^2\hat H_2
-\hat H_2\hat \rho^2\hat H_1
\big)\hat \rho.
\label{5'}
\ee
The RHS of (\ref{5'}) is Hermitian if $\hat \rho$, $\hat H_1$ and
$\hat H_2$ are Hermitian. This explains the choice of {\it real\/}
$z_5$. 
For $\hat \rho=\hat \rho^2$ (\ref{5'}) vanishes. 
Assuming that $\rho(0)=\rho(0)^*$ we find that all higher
derivatives are also Hermitian.
The formal solution 
\be
{}&{}&\rho_a(t)=\sum_{n=0}^\infty
\frac{t^n}{n!}\nonumber\\
&{}&\times
\underbrace{\{\dots\{\{}_n\rho_a(0),H_1,H_2,S_1,S_2\}
,H_1,H_2,S_1,S_2\}\nonumber\\
&{}&
\phantom{
\times
\underbrace{\{\dots\{\{}_n\rho_a(0),H_1,H_2,S_1,S_2\}
}
\dots,H_1,H_2,S_1,S_2\}
\label{szereg'}
\ee
satisfies $\rho(t)=\rho(t)^*$ if $\rho(0)=\rho(0)^*$.
Using the same argument as for $n=3$ we conclude that the
spectrum of self-adjoint and Hilbert-Schmidt solutions of
(\ref{5}) is conserved.

\section{$N$-particle extensions of 1-particle almost-Lie-Poisson dynamics}

An extension of dynamics from 1 to $N$ particles
is a delicate problem. Careful analysis shows that the Lie-Nambu
duality holding for the 3-brackets leads to generalizations
which behave differently from the viewpoint of $N$-particle
extensions. The Poisson dynamics based on
$\{\,\cdot\,,\,\cdot\,\}_S$ for $S=C_2/2$ is the most regular
one. An inclusion of nonlinear Hamiltonian functions $H$ does not lead to 
difficulties with independent evolutions of separated systems. 
This fact was proved by Polchinski (1991)
and Jordan (1993) and recently generalized
by myself to those nonlinear theories which do not possess
Hamiltonian {\it functions\/} but only Hamiltonian {\it
operators\/} (Czachor, 1997b).  
In spite of this, the view that {\it any\/} nonlinear
generalization of a Schr\"odinger dynamics leads to problems with
causality is quite popular. Nonlinear quantum mechanics based on
$\{\,\cdot\,,\,\cdot\,\}_{-H}$ leads to a new kind of nonlocal
phenomenon. 
This effect, typical of
mixed states, is analogous to the threshold phenomena
discussed by Goldin and Svetlichny (1994) for pure states. 

\subsection{$N$-particle ``metric" tensors}

Let $g_{a_1\dots a_n}$ be a 1-particle metric tensor. The
$N$-particle tensor is defined by 
\be
g^N_{a_1\dots a_n}=g_{a^1_1\dots a^1_n}\dots g_{a^N_1\dots
a^N_n}. \label{g^N}
\ee
The indices at the LHS of (\ref{g^N}) are the $N$-particle ones:
$a_1=a^1_1\dots a^N_1$, etc. The $N$-particle $n$-bracket is
defined by 
\be
\{A_1,\dots, A_n\}^N=\Omega^N_{a_1\dots a_n}
\frac{\delta A_1}{\delta \rho^N_{a_1}}\dots
\frac{\delta A_n}{\delta \rho^N_{a_n}},\label{N-n}
\ee
where $\rho^N_a=\rho_{a^1\dots a^N}$
\be
\Omega^N_{a_1\dots a_n}=(n-1)!g^N_{[a_1\dots a_n]}.
\ee
For identical particles
(bosons and fermions) $\rho_{a^1\dots a^N}=\rho_{(a^1\dots
a^N)}$ with $(\dots)$ denoting symmetrization. A distinction
between fermions and bosons can be seen at the ``spinor index"
level: 
\be
\rho_{a^1\dots a^N}&=&\rho_{[\alpha^1|\alpha'^1|\dots
|\alpha^N]\alpha'^N}= 
\rho_{\alpha^1[\alpha'^1|\dots |\alpha^N|\alpha'^N]}
\,{\rm (fermions)},\nonumber\\
\rho_{a^1\dots a^N}&=&\rho_{(\alpha^1|\alpha'^1|\dots
|\alpha^N)\alpha'^N}= 
\rho_{\alpha^1(\alpha'^1|\dots |\alpha^N|\alpha'^N)}
\,{(\rm bosons)}.\nonumber
\ee
For the 3-bracket the annihilation property implies the
important identity 
\be
{}&{}&
\Omega^N_{abc}\omega^{b_1}\dots
\omega^{b_{k-1}}\omega^{b_{k+1}}\dots \omega^{b_N}\nonumber\\
&{}&=
g_{a_1c_1}\dots g_{a_{k-1}c_{k-1}}\Omega_{a_kb_kc_k}
g_{a_{k+1}c_{k+1}}\dots g_{a_Nc_N},\label{identity}
\ee
where $\Omega_{a_kb_kc_k}$ are the 1-particle structure
constants of the $k$-th particle. 

\subsection{Extension of Hamiltonians} 

The 3-bracket dynamics is Lie-Poisson if $S=C_2/2$ and a
Hamiltonian function $H$ exists. There exist nonlinear
Liouville-von Neumann equations that possess
Hamiltonian operators of the form $\hat H(\rho)=
\hat H_1(\rho)+\hat H_2(\rho)$ where $\Tr \hat H_2(\rho)\hat
\rho=0$. Such equations do not possess a Hamiltonian function 
$\Tr \hat H(\rho)\hat \rho$ but often are of physical
interest (e.g. starting with non-linearizable Doebner-Goldin 
Schr\"odinger equations (Doebner and Goldin, 1996) 
one arrives at this class of mixed state equations (Czachor, 1997b)). 
Equations that can be written as 
\be
\dot \rho_a=
-i\Omega_{abc}H^b(\rho)\rho^c,\label{aLP}
\ee
although no $H$ satisfying $H^b= \frac{\delta H}{\delta\rho_b}$
exists, will be called almost-Lie-Poisson. Assume we have $N$
(not necessarily identical) particles that do not interact with one
another (but can interact with something else and do not have to
be free). Each of them satisfies a 1-particle equation
(\ref{aLP}) with some $H$. 
We define the $N$-particle extension
of (\ref{aLP}) by
\be
\dot \rho^N_a=
-i\Omega^N_{abc}H^{b}(\rho^N)\rho^{Nc},\label{NaLP}
\ee
where 
\be
{}&{}&
H^{b}(\rho^N)\nonumber\\
&{}&
=H_1^{b_1}(\rho_{(1)})\omega^{b_2}\dots \omega^{b_N}
+\dots +
\omega^{b_1}\dots \omega^{b_{N-1}} H_N^{b_{N}}(\rho_{(N)}).\label{NH}
\ee
The reduced density matrix $\rho_{(k)}$ is defined by 
\be
\rho_{(k) a_k}=\omega^{a_1}\dots
\omega^{a_{k-1}}\omega^{a_{k+1}}\dots \omega^{a_N}
\rho_{a_1\dots a_k\dots a_N}.
\ee
There are two motivations for (\ref{NH}). First, if the $k$-th
particle is described by a Hamiltonian function
$H_k(\rho)=H_k(\rho_{(k)})$ then (\ref{NH}) is just a
consequence of the chain rule for functional derivatives.
The second motivation is (\ref{identity}). Indeed, applying (\ref{identity})
to (\ref{NaLP}) we obtain 
\be
{}&{}&
i\dot \rho_{a_1\dots a_N}
=
\Omega_{a_1b_1c_1}H_1^{b_1}(\rho_{(1)})\rho{^{c_1}}_{a_2\dots a_N}
\nonumber\\
&{}&\phantom{xxxxxxx}+
\dots 
+
\Omega_{a_Nb_Nc_N}H_N^{b_N}(\rho_{(N)})\rho_{a_1\dots
a_{N-1}}{^{c_N}}. \label{2nd motiv}
\ee
Transvecting both sides of (\ref{2nd motiv}) with
$$
\omega^{a_1}\dots \omega^{a_{k-1}}\omega^{a_{k+1}}\dots
\omega^{a_N}
$$ 
and using $\omega^a\Omega_{abc}=0$ we get 
\be
\dot \rho_{(k)a}=
-i\Omega_{a_kb_kc_k}H_k^{b_k}(\rho_{(k)})\rho_{(k)}^{c_k}.\label{(k)aLP}
\ee
Both sides (\ref{(k)aLP}) depend only on objects which are
intrinsic to the $k$-th subsystem. It follows that the reduced
density matrix of this subsystem ``does not see" the
other noninteracting systems. {\it The observers in the other
subsystems have no possibility of influencing the dynamics of
the $k$-th one by any kind of modification of the Hamiltonians
in the other separated systems\/}. In particular, they cannot
influence any observable quantity in the $k$-th subsystem by
different choices of measurements in their ``own" subsystems.  
This explicitly contradicts the popular claim that any nonlinear
dynamics must imply faster than light influences between
separated systems. 

Denoting the dynamics of the
$N$-particle system by $\phi^t_N$, the one corresponding to the
$k$-th subsystem by $\phi^t_k$, and by $\Tr_{N-k}$ the partial
trace which reduces the dynamics from the composite system to
the $k$-th one, we get an important separability condition 
\be
\Tr_{N-k}\circ\phi^t_N=\phi^t_k\circ\Tr_{N-k} \label{(17)}
\ee
characteristic of the Lie-Poisson dynamics. The dynamics
satisfying (\ref{(17)}) and (\ref{(k)aLP}) can be termed {\it
completely separable\/}. 

\subsection{Examples of completely separable extensions}

The method of extension given by (\ref{NH}) applies to any
equation whose 1-particle Hamiltonian operator can be written as a
function of the particle's {\it density matrix\/}. This 
applies also to pure state (Schr\"odinger) equations. To see how
this works for nonlinear Schr\"odinger equations consider the examples.
A Hamiltonian operator consists of two parts: 
The linear part $\hat H_L(x)=\hat H_{\rm kinetic} + V(x)$ 
and a nonlinear $\hat H_{NL}=\hat H(\psi,\bar
\psi;x)$. To apply the above method we have to be able to write 
$\hat H(\psi,\bar \psi;x)$ as $\hat H(\rho;x)$.

\medskip
\noindent
a) {\it ``Nonlinear Schr\"odinger equation"\/}.
\begin{eqnarray}
\hat H(\psi,\bar \psi;x)=
|\psi(x)|^2\to \hat H(\rho;x)=\rho(x,x)\nonumber
\end{eqnarray}
b) {\it Bia{\l}ynicki-Birula--Mycielski equation\/}
(Bia{\l}ynicki-Birula--Mycielski, 1976).
\begin{eqnarray}
\hat H(\psi,\bar \psi;x)=
\ln\big(|\psi(x)|^2\big)\to \hat H(\rho;x)=\ln\rho(x,x)\nonumber
\end{eqnarray}
Obviously in the same way one can treat any equation with
nonlinearities given by some function $H(|\psi(x)|)$.

\bigskip\noindent
c) {\it Haag-Bannier equation\/} (Haag and Bannier, 1978)
\be
\hat H(\psi,\bar \psi;x)&=&
\vec A(x)
\frac{\bar \psi(x)\vec \nabla_x \psi(x) -\psi(x)\vec \nabla_x
\bar \psi(x)}{2i|\psi(x)|^2}\to \nonumber\\
\hat H(\rho;x) &=&
\vec A(x)
\frac{\int dz\delta(x-z)\vec \nabla_x[\rho(x,z)-\rho(z,x)]}
{2i\rho(x,x)}
\ee
d) {\it Doebner-Goldin equations\/} (Doebner and Goldin, 1996;
Nattermann, 1997). 
There are five nonlinear terms
denoted by $R_k$:  
\begin{eqnarray}
R_1(\psi,\bar \psi;x)&=&
\frac{1}{2i}
\frac{
\bar\psi(x)\Delta_x \psi(x)
-
\psi(x)\Delta_x \bar\psi(x)}
{|\psi(x)|^2 }\to\nonumber\\
R_1(\rho;x)&=&
\frac{1}{2i}
\frac{\int dz\delta(x-z)
\Delta_x \big[\rho(x,z)
-\rho(z,x)\big]}
{\rho(x,x)},\nonumber\\
R_2(\psi,\bar \psi;x)&=&
\frac{\Delta_x|\psi(x)|^2}{|\psi(x)|^2}
\to R_2(\rho;x)=
\frac{\Delta_x\rho(x,x)}{\rho(x,x)}\nonumber\\
R_3(\psi,\bar \psi;x)&=&
\frac{1}{(2i)^2}
\frac{
\bigl[
\bar \psi(x)\vec \nabla_x \psi(x)
-
\psi(x)\vec \nabla_x \bar \psi(x) \bigr]^2}
{|\psi(x)|^4}\to \nonumber\\
R_3(\rho;x)&=&
\frac{1}{(2i)^2}
\frac{\big(
\int dz\delta(x-z)
\vec \nabla_x \big[\rho(x,z)
-\rho(z,x)\big]\big)^2}
{\rho(x,x)^2}\nonumber\\
R_4(\psi,\bar \psi;x)&=&
\frac{1}{2i}
\frac{
\bigl[
\bar \psi(x)\vec \nabla_x \psi(x)
-
\psi(x)\vec \nabla_x\bar \psi(x)
\bigr]\cdot\vec \nabla_x |\psi(x)|^2}
{|\psi(x)|^4}\to\nonumber\\
R_4(\rho;x)&=&
\frac{1}{2i}
\frac{\int dz\delta(x-z)
\vec \nabla_x \big[\rho(x,z)
-\rho(z,x)\big]\cdot \vec \nabla_x\rho(x,x)}
{\rho(x,x)^2},\nonumber\\
R_5(\psi,\bar \psi;x)&=&
\frac{
\bigl[
\vec \nabla_x |\psi(x)|^2\bigr]^2}
{|\psi(x)|^4}
\to R_5(\rho;x)=
\frac{
\bigl[
\vec \nabla_x \rho(x,x)\bigr]^2}
{\rho(x,x)^2}\nonumber
\end{eqnarray}
e) {\it Twarock equation\/} on $S^1$ (Twarock, 1997)
\begin{eqnarray}
\hat H(\psi,\bar \psi;x)&=&
\frac{\psi(x)'' \overline{\psi(x)'}
-
\overline{\psi(x)''}\psi(x)'}
{\psi(x) \overline{\psi(x)'}
-
\overline{\psi(x)} \psi(x)'}\to \nonumber\\
\hat H(\rho;x) &=&
\frac{
[\int dy\delta(x-y)\partial_x^2\rho(x,y)]
[\int dz\delta(x-z)\partial_x^2\rho(z,x)] - c.c.}
{\rho(x,x)\int dy\delta(x-y)\partial_x\rho(x,y) - c.c.}\nonumber
\end{eqnarray}
f) $(n,n)$-{\it homogeneous nonlinearities\/}. Denote by
$D$ a differential operator involving
arbitrary mixed partial derivatives up to order $k$. 
Consider a  real function
$H(\psi)=F\big(D\psi(x)\big)$, which is 
$(n,n)$-homogeneous i.e. satisfies $H(\lambda\psi)=\lambda^n
\bar \lambda^n H(\psi)$. We first write 
\begin{eqnarray}
F\big(D\psi(x)\big)=
\frac{F\big(\overline{\psi(x)}D\psi(x)\big)}
{|\psi(x)|^{2n}}\nonumber
\end{eqnarray}
and then apply the tricks used for the Haag--Bannier, 
Doebner--Goldin and Twarock
terms. Obviously any reasonable function of such
$(n,n)$-homogeneous expressions with different $n$'s will do as well.

Let us now concentrate on the simplest case with
$H_k(\rho;x)=H_k\big(\rho(x,x)\big)$ and just two particles. 
The 2-particle extension of the nonlinear part of the
Hamiltonian is 
\be
{}&{}&
\hat H_1\big(\rho_{(1)}(x_1,x_1)\big) + \hat
H_2\big(\rho_{(2)}(x_2,x_2)\big) \nonumber\\
&{}&=
H_1({\textstyle\int} dy \rho(x_1,y,x_1,y)+
H_2({\textstyle\int} dy \rho(y,x_2,y,x_2).\label{1+2}
\ee
If the 2-particle state is pure $\rho(x_1,x_2,x'_1,x'_2)=
\Psi(x_1,x_2)\Psi^*(x'_1,x'_2)$ the RHS becomes 
\be
H_1\big({\textstyle\int }dy |\Psi(x_1,y)|^2\big)+
H_2\big({\textstyle\int } dy |\Psi(y,x_2)|^2\big)
\ee
and reduces to   
\be
H_1\big(|\psi(x_1)|^2\big)+H_2\big(|\phi(x_2)|^2\big)
\ee
on product states $\Psi(x_1,x_2)=\psi(x_1)\phi(x_2)$. 

It seems that an example that cannot be treated in this
way is the Kostin equation (Kostin, 1972) involving nonlinearity $\ln
\big(\psi(x)/\overline{\psi(x)}\big)$. 

A reader may have noticed that the above reasoning involves two
``heretic" elements. First of all the 2-particle extension of
dynamics for {\it nonfactorizable\/} (entangled) states leads to
integro-differential equations. Such equations are typically rejected in
nonlinear quantum mechanics literature as {\it nonlocal\/}. The
construction presented above shows that it is in fact just the
opposite. The requirement of locality (complete separability)
leads us to {\it appropriate\/} integral terms and precisely
because of these terms the subsystems can be completely isolated
from one another. Second, {\it all\/} nonlinearities of the form
$F\big(|\psi(x)|^2\big)$ are acceptable. This is in an apparent
contradiction with the well known result of Bia{\l}ynicki-Birula and
Mycielski who used the separability criterion to {\it derive\/} the
logarithmic nonlinearity. However, they assumed that the
2-particle extension has to be $F\big(|\Psi(x_1,x_2)|^2\big)$
and, with the condition 
\be
F\big(|\psi(x_1)\phi(x_2)|^2\big)=
F\big(|\psi(x_1)|^2\big)
+
F\big(|\phi(x_2)|^2\big),\label{F+F}
\ee
they found that only $F\sim \ln$ is acceptable. One of the
obvious drawback of such extensions is that they do not tell us
what to do if the systems are noninteracting but correlated 
and $\Psi(x_1,x_2)$ does not factorize. In such a case {\it
local\/} probability densities are obtained by integrating
out the coordinates of the remaining particles and it is quite
logical that such expressions occur in the $N$-particle
``correctly extended" Hamiltonians we discussed. 

\subsection{Problem of complete positivity}

A subsystem described by $\rho_{(k)}$ can be 
embedded into a composite one described by $\rho^N$ in a way
guaranteeing a consistency of (\ref{NaLP}) and (\ref{(k)aLP}).
The dynamics of $\rho_{(k)}$ is independent of $N$. In addition,
since both $\rho_{(k)}$ and $\rho^N$ satisfy the 3-bracket
Lie-Poisson equation, the extension procedure preserves
positivity of dynamics at both subsystem and composite
system levels. A dynamics that has these properties is typically
associated with the notion of a {\it completely\/} positive map,
provided the maps are {\it linear\/}. 

In mathematical literature the notion of complete positivity is 
generalized to nonlinear maps in a way that can be translated
to our context as follows (Ando and Choi, 1986; Arveson, 1987;
Majewski, 1990; Alicki and Majewski, 1990). One
takes a positive map 
\begin{eqnarray}
\phi_1^t(a)= a(t),\quad \phi_1^t:\, {\cal A}\to  {\cal A}
\end{eqnarray}
where ${\cal A}$ is a unital $C^*$-algebra. In our case
$a=\rho_{(k)}$ and $\phi_1^t(a)= \rho_{(k)}(t)$. Assume for
simplicity that the dimension of the $k$-th system is finite. 
In the next step one considers a density matrix $\rho^N(0)$
of a bigger system consisting of the original one plus a system which
has a  finite number $m$ of degrees of freedom. Writing 
\be
\rho^N(0)&=&
\sum_{r,r',s,s'}\rho^N(0)_{rr'ss'}|r\rangle\langle r'|
\otimes
|s\rangle\langle s'|\nonumber\\
&=&
\sum_{s,s'}a_{ss'}
\otimes
|s\rangle\langle s'|\label{rr'ss'0}
\ee
we can represent $\rho^N(0)$ by the matrix 
\be
\left(
\begin{array}{ccc}
a_{11} & \dots & a_{1m} \\
\vdots & \vdots & \vdots\\
a_{m1} & \dots & a_{mm}
\end{array}
\right)\label{akl0}
\ee
whose entries are elements of $\cal A$. The (nonlinear) map $\phi_1^t$ is
said to be completely positive if the matrix 
\be
\left(
\begin{array}{ccc}
\phi_1^t(a_{11}) & \dots & \phi_{1}^t(a_{1m}) \\
\vdots & \vdots & \vdots\\
\phi_{1}^t(a_{m1}) & \dots & \phi_{1}^t(a_{mm})
\end{array}
\right)\label{phi12(t)}
\ee
is positive for any $m$. This is equivalent to the positivity of
\be
\tilde \phi^t\big(\rho^N(0)\big)
=\sum_{s,s'}\phi^t_1(a_{ss'})
\otimes
|s\rangle\langle s'|.\label{rr'ss't}
\ee
However, for nonlinear $\phi^t_1$ the explicit form of
(\ref{rr'ss't}) for $t>0$ is different for differenet choices of
bases $\{|s\rangle\}$, which is unphysical unless there exists a
superselection rule distinguishing a particular basis. In the
generic case no such distinguished basis exists. Therefore a {\it basis
independent\/} extension from 1 to more particles cannot have
the forms (\ref{phi12(t)}) and (\ref{rr'ss't}). 
And, indeed, the dynamics following from the Lie-Poisson extension 
discussed above does not coincide with (\ref{rr'ss't}). This was
shown by an explicit calculation in Czachor and Kuna (1997b)  but could be
inferred also from the basis-independence of the $N$-particle
extension. It must be stressed that the dynamics (\ref{rr'ss't})
is the one that was used by Gisin (1989) in his discussion of
unphysical influences between separated systems. 

\section{Nonlocal properties of $N$-particle extensions for the
dual Poisson structure $\{\,\cdot\,,\,\cdot\,\}_{-H}$}

The regularity of the $N$-particle extensions typical of an
almost-Lie-Poisson dynamics is lost when one considers the dual
Poisson structure $\{\,\cdot\,,S\}_{-H}$ with $S$ being a higher
order Casimir invariant. To explicitly see the kind of
difficulties one may encounter consider the 2-particle equation 
\be
i\dot \rho_{a_1a_2}=\{\rho_{a_1a_2},C_3/3\}_{-H}.
\ee
General properties of the 3-bracket dynamics imply that
$C_n(\hat \rho)$ are conserved for any 
natural $n$, where $\hat \rho$ is
the 2-particle density matrix. Also $C_1(\hat \rho_{(1)})$ is a
constant of motion. However
\be
i\dot C_2(\hat \rho_{(1)})
&=&
2\Tr_1\Big([\Tr_2(\hat \rho^2),\Tr_2(\hat \rho)] \hat H_1\Big),\label{"BB"}
\ee
where the indices 1 and 2 correspond to the two subsystems and
we have assumed the standard 2-particle extension of the
(linear) Hamiltonian. Although we do not have much control over the
behavior of the eigenvalues $p_j$ of the reduced density matrix $\hat
\rho_{(1)}$, we can infer that $\sum_jp_j$ is constant whereas 
$\sum_jp^2_j$ is, in general, time dependent. Let us note that average
energies of the two subsystems are separately conserved. This
follows from the general property of the 3-bracket: For 
$H(\hat \rho)=H_1(\hat \rho_{(1)})+H_2(\hat
\rho_{(2)})$
\be
\{H_1(\hat \rho_{(1)}), H(\hat \rho), S(\hat \rho)\}=0
\ee
for any $S$ (Czachor, 1997a). Therefore the probabilities $p_j$ can
be made time dependent without making the two subsystems
interact with each other and without changing energies of the
subsystems just by modifying the overall entropy of the composite
system. So the change of entropy, say, by $C_2\to C_2 +
\epsilon C_3$ at the global level leads to the modification of
the local subsystems. Such a modification will not occur if 
\be
[\Tr_2(\hat \rho^2),\Tr_2(\hat \rho)]=0
\ee
which holds for a pure-state $\hat \rho$, or $\hat \rho=\hat
\rho_{(1)}\otimes \hat \rho_{(2)}$. Still, strong correlations can also
help since reduced density matrices occuring in a singlet state
are proportional to unit matrices and the commutator vanishes. 
Systems described by entropies other than $C_2/2$ possess some
kind of overall identity which is lost when it is physically
meaningful to discuss their subsystems separately.  
This effect deserves a name. The fact that the subsystems ``feel"
that the total entropy (information) undegoes a change from
$C_2$ to $C_2 + \epsilon C_3$ although apparently ``nothing
happened" (no energy has been transfered between the neighboring
subsystems) resembles the influence the Big Brother from
G.~Orwell's ``1984" exerted on the inhabitants of Oceania ``just
by watching them". 

It is not immediatly clear that the ``Big Brother effect" is entirely
unphysical. Its interpretation is obscured by our lack of
understanding of the physical role played by the entropies $C_n$
in the multiple-bracket scheme. 
It may be relevant to mention
that $C_2$ is characteristic of R\'enyi 2-entropy which is the
only $\alpha$-entropy that characterizes a system whose gain of
information is 0 for all probability distributions. The analysis of this
problem was given by A.~R\'enyi (1960). Although this is
the {\it first\/} paper where the notion of $\alpha$-entropies was
introduced, it does not seem to be known to the majority of experts in
quantum mechanical information theory. The work typically
quoted in the literature is R\'enyi (1961) where R\'enyi already departed
from the natural definition of the information gain in favor of
a ``decrease of uncertainty". This latter modification was
motivated by the problem with the vanishing gain for $\alpha=2$.

A class of physical systems whose identity as a whole is
associated with the way 
their entropy (or information) behaves are {\it living\/}
organisms. 
Similarly, statistical properties of {\it
societies\/} have dynamical properties strongly depending on
information, and their dynamics cannot be regarded as a simple
sum of individual activities.  
The fact that a possibility of gaining information can be
formally related, via $S$, to nonlinearity of evolution
resembles a similar phenomenon mentioned by Wigner (1967) in the context
of the measurement problem (``paradox of a friend").
Whether such phenomena
are in any way related to the 3-bracket dynamics is at the
moment a matter of pure speculation.

\section{Quantization of classical Nambu dynamics?}

The $(2n+1)$-bracket can be regarded as a nonlinear quantization
of a classical $(n+1)$-bracket with $n$ classical Hamiltonian
functions $H_1,\dots ,H_n$. Indeed, the Liouville-von Neumann
equation is characterized by {\it one\/} Hamiltonian operator
$\hat H_1$, obtained by a quantization of a classical
Hamiltonian function $H_1$. The requirement of linearity of
evolution combined with the 3-bracket dynamics leads to the
choice of $S=C_2/2$. Having $n$ Hamiltonian functions $H_k$ we
can obtain $n$ Hamiltonian operators $\hat H_k$ after {\it some\/}
quantization procedure (say, $p\to -i\hbar\nabla_x$, etc.). 
Representing the operators by kernels $H_k^{a_k}$, $k=1,\dots,n$, we can
consider the $(2n+1)$-bracket equation 
\be
\dot \rho_a=z_{2n+1}\Omega_{aa_1\dots a_nb_1\dots b_n}
H_1^{a_1}\dots H_n^{a_n}
\frac{\delta S_1}{\delta \rho_{b_1}}\dots 
\frac{\delta S_n}{\delta \rho_{b_n}}.
\ee
For $n>1$ the equation is always nonlinear and its RHS vanishes on pure
states. A self-adjoint Hilbert-Schmidt solution of the
``quantized Nambu dynamics" may be interpreted as a density matrix
because spectrum of the solution is conserved. Had we started
with the {\it linear\/} equation, which could be obtained by
taking the $(n+2)$-bracket  
\be
\dot \rho_a=z_{n+2}\Omega_{aa_1\dots a_na_{n+1}}
H_1^{a_1}\dots H_n^{a_n}\rho^{a_{n+1}},
\ee
we would have obtained a dynamics which would not, in general,
conserve $\Tr(\hat \rho^m)$ for $m>1$ and there would be no
guarantee that positivity of $\hat \rho$ is conserved. 

This kind of nonlinear quantization differs from the procedure
discussed in Takhtajan (1994) which was based on an
$n$-bracket obtained by an antisymmetrization of a product of
$n$ oparators, or the Zariski product quantization proposed in
Ditto {\it et al.\/} (1997). Also all operator
expressions involving an odd number of operator kernels, if
described within our approach, must be excluded because the
``metric" tensor used for the generalized structure constants
would have to have an even number of indices, but such structure
constants vanish (3-bracket involves antisymmetrization of two
operators, 5-bracket antisymmetrizes four operators, etc.). 
The quantization proposed originally by Nambu (1973) (cf. Garcia
Sucre and K\'alnay, 1975) is
therefore also inequivalent to our formulation.

\section{What next?}

The formalism presented in this work is at a very preliminary
stage of development. The main problem is how to solve the
nonlinear density matrix equtions and how to extend the approach
to a fully relativistic theory. Both questions are highly
nontrivial. The equations of the form $i\dot \rho=[\hat
H, \rho^n]$ bear some formal similarity to the Nahm
equations studied in the $SU(2)$ monopole theory (Hitchin, 1983).
Some recently developed techniques of solving
matrix equations by a noncommutative version of a Darboux
transform (Leble and Zaitsev, 1997; Leble, 1997) 
may prove useful in this context.
A natural
candidate for a relativistic multiple-bracket formalism is the off-shell
proper-time formulation. Some preliminary indefinite-metric results
can be found in Czachor and Kuna (1997a), and a work on
positive-metric generalization of the Bargmann-Wigner off-shell
equations (Czachor, 1997c) is in progress. The characteristic
function off-shell approach developed recently by Naudts (1998) seems
especially suited for this kind of generalization.
A separate problem is the behavior of eigenvalues of reduced
density matrices in the $\{\,\cdot\,,\,\cdot\,\}_{-H}$ scheme.
These eigenvalues are in general time dependent and therefore
the question of their interpretation is still unclear.

\bigskip
This work was written  during my stays in Arnold
Sommerfeld Institute in Clausthal  and in Centrum
Leo Apostel in Brussels. Financial supports from DAAD
and the Polish-Flemish grant 007 are gratefully acknowledged.

\end{document}